\documentclass[sigconf,authorversion,screen]{acmart}
\AtBeginDocument{%
  \providecommand\BibTeX{{%
    \normalfont B\kern-0.5em{\scshape i\kern-0.25em b}\kern-0.8em\TeX}}}

\copyrightyear{2022}
\acmYear{2022}
\setcopyright{rightsretained}
\acmConference[ETRA '22]{2022 Symposium on Eye Tracking Research and Applications}{June 8--11, 2022}{Seattle, WA, USA}
\acmBooktitle{2022 Symposium on Eye Tracking Research and Applications (ETRA '22), June 8--11, 2022, Seattle, WA, USA}
\acmDOI{10.1145/3517031.3531628}
\acmISBN{978-1-4503-9252-5/22/06}

\begin{document}
\title{Interaction Design of Dwell Selection Toward Gaze-based AR/VR Interaction}

\author{Toshiya Isomoto}
\email{isomoto@iplab.cs.tsukuba.ac.jp} %
\affiliation{
\institution{University of Tsukuba}
\city{Tsukuba}
\state{Ibaraki}
\country{JAPAN}
}
\author{Shota Yamanaka}
\email{syamanak@yahoo-corp.jp} %
\affiliation{
\institution{Yahoo Japan Corporation}
\city{Chiyoda}
\state{Tokyo}
\country{JAPAN}
}
\author{Buntarou Shizuki}
\email{shizuki@cs.tsukuba.ac.jp} %
\affiliation{
\institution{University of Tsukuba}
\city{Tsukuba}
\state{Ibaraki}
\country{JAPAN}
}

\renewcommand{\shortauthors}{Isomoto et al.}

\begin{abstract}
In this paper, we first position the current dwell selection among gaze-based interactions and its advantages against head-gaze selection, which is the mainstream interface for HMDs. 
Next, we show how dwell selection and head-gaze selection are used in an actual interaction situation. 
By comparing these two selection methods, we describe the potential of dwell selection as an essential AR/VR interaction.
\end{abstract}



\maketitle

\section{Introduction}
Dwell selection has been the mainstream method since the gaze-based interface was proposed~\cite{Jacob:1990:YLY:97243.97246}.
Although numerous researchers have put their efforts into dwell selection to solve the critical issue of unintended selection, which is referred to as the Midas-touch problem~\cite{Jacob:1990:YLY:97243.97246}, dwell selection still faces this issue.
Dwell selection for a target requiring much time to understand its content (e.g., a target consisting of a text and image) is not always achievable~\cite{speedaccuracyPETMEI11}, particularly in contrast to the successful achievement of dwell selection for a simple target.
When selecting such non-simple targets, it is difficult to predict whether the user intends to select a target or not with only the time threshold as the required time varies depending on the content.
Although many studies have pointed out the advantages of dwell selection highly appreciated for AR/VR interaction, such as ``easy to use'' and ``straightforward,'' they also reported that users generally dislike this selection method as a result of the frequent occurrence of Midas-touch problems (e.g.,~\cite{GlanceVergeSUI21}).

Many selection methods that uses a combination of gaze and other modalities have been proposed to solve the Midas-touch problem. 
Specifically, with the recent development of an HMD with a built in eye-tracker, a selection method using a combination of gaze and head-rotation (i.e., the direction of the head) has been researched. 
This method is referred to as head-gaze selection, in which users can select a target by dwelling on the button while their head is directed toward the target (e.g.,~\cite{confirmationbuttonactigazeUIST15,GlanceableAR}). 
Given that no interaction is performed by dwelling on a target, a few Midas-touch problems occur.
Even though head-gaze selection has better results than current dwell selection against the occurrences of the Midas-touch problem, the potential of dwell selection, in terms of easy and straightforward selection, remains attractive. 
 
In this paper, we show an interaction design of dwell selection toward gaze-based AR/VR interaction and describe the potential of dwell selection.
\begin{figure*}
    \centering
    \includegraphics{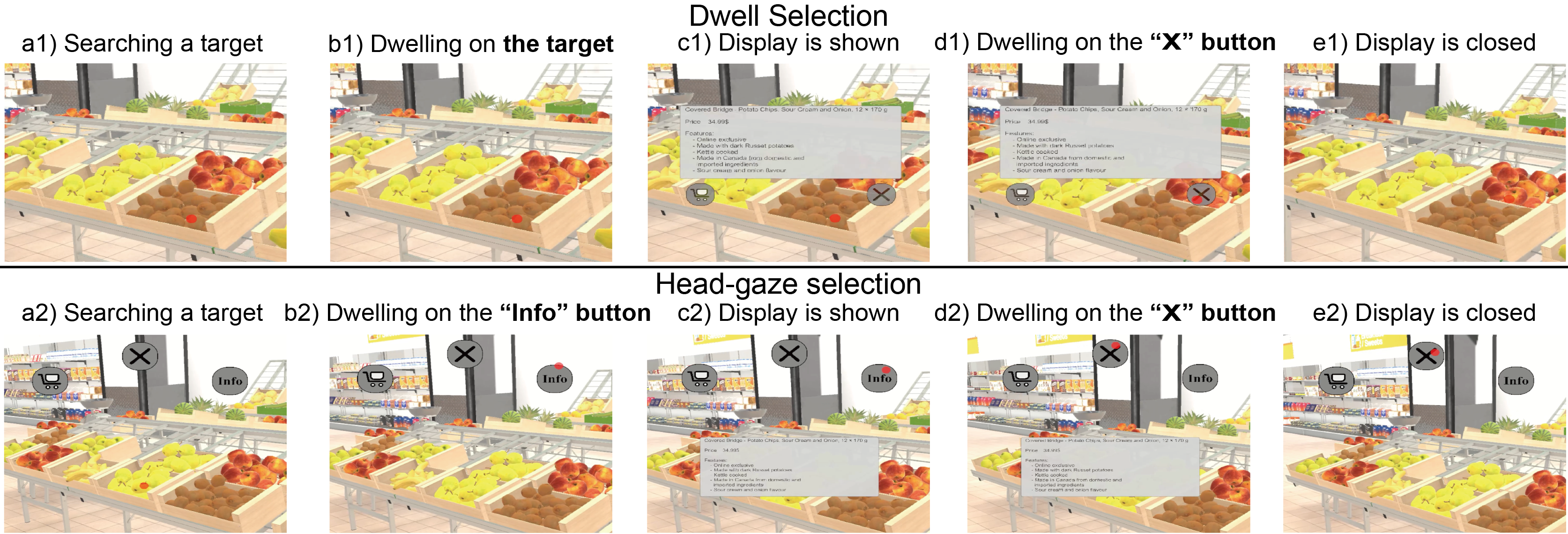}
    \caption{Schema of two selection methods in a shopping situation. These images are made by cropping the original image almost into the user's field of view (FOV). (Top) The user of dwell selection can show information and corresponding buttons by dwelling on a target (a1--c1) and close them by dwelling on the ``X'' button (d1 and e1). (Bottom) The user of head-gaze selection can show information by dwelling on the ``info'' button positioned at the outermost of the FOV (a2--c2) and close it by dwelling on the ``X'' button positioned at the outermost of the FOV (d2 and e2). The red point shows the gaze position of the user.}
    \label{fig:methods}
\end{figure*}

\section{Interaction Design of Dwell Selection}
We show the use of dwell selection and head-gaze selection and compare the two methods to show the potential of dwell selection to be an essential AR/VR interaction.

With dwell selection (Figure~\ref{fig:methods}, top), a user can display food information by dwelling on it and close the display by dwelling on a button. 
In the example, there are many possibilities where the Midas-touch problem can occur when selecting food and beverages because their content, such as price and nutrition, depends on each item.
Accordingly, the user must take much time to understand the content, and there are possibilities for causing the Midas-touch problem.
Occurrences of the Midas-touch problem could irritate users due to the frequent unwanted display of information.
Recently, a few studies have begun developing the robustness of dwell selection against the Midas-touch problem with machine-learning based intent prediction (e.g.,~\cite{intentVRcontrollerBrendanETRA21,intentBednarikETRA12}). 
Although such dwell selection has a great possibility of preventing the Midas-touch problem, improving the prediction performance is still necessary for robust selection. 
We should use a dwell selection with intent prediction for non-simple targets due to the necessity of robustness against the Midas-touch problem and use a dwell selection with only a time threshold for a simple target because robustness has already been achieved. 

With head-gaze selection (Figure~\ref{fig:methods}, bottom), the contrast to dwell selection are the separation of target and buttons and the buttons' position. 
Most buttons are consistently positioned at the outermost of the user's field of view (FOV) to avoid the Midas-touch problem. 
That is, the selection is performed by first moving the gaze onto a button positioned at the outermost of the FOV and then dwelling on it. 
In Figure~\ref{fig:methods}, selecting a button shows and closes the information mapped into the button. 
Even though head-gaze selection requires additional eye movement and buttons, it achieves high usability~\cite{GlanceableAR} and thus has been the promised interface for an HMD.

Having no need for additional eye movement increases dwell selection's potential.
For example, one important potential is smaller physical demand. 
Although head-gaze selection requiring additional eye movement and buttons achieves high usability~\cite{GlanceableAR}, moving the eyes toward the outermost of the FOV could be physically demanding for users~\cite{vorLudwigUIST19}. 
With dwell selection, a user can select a target just by dwelling on the target (Figure~\ref{fig:methods}b1), followed by searching for a target (Figure~\ref{fig:methods}a1) without additional eye movement. 
Another potential is the ability of free interface design.
The outermost of the FOV that may affect the user's attention least is useful for stacking notifications and displaying status bars in the same manner as in desktop interfaces.
Also, dwell selection is helpful for combined interactions with the mouse, touch, and voice.
These combined interactions use dwell selection to determine the user's intent in the same manner as the principle of dwell selection of ``what the user looks at is what the user wants.''
Thus, dwell selection with precise intent prediction will be essential for a gaze-assisted interface. 

\section{Conclusion}
We showed an interaction design of dwell selection toward gaze-based AR/VR interaction, describe the potentials of dwell selection, and pointed out two factors requiring dwell selection to be such an interface.
One is improving the user's intent prediction for dwell selection without any additional eye movement and buttons, and the other is investigating the performance of dwell selection with a target that requires much time to understand its content.
We believe that the proposed dwell selection will become an essential gaze-based AR/VR interaction.

\bibliographystyle{ACM-Reference-Format}
\bibliography{sample-base}

\end{document}